\documentclass[12pt]{iopart}

\usepackage{graphicx}
\graphicspath{ {Images_SMS/} }
\usepackage{float}

\begin{document}

\title[Visualization of Stress Waves via Air-Coupled Acoustic Emission Sensors]{Visualization of Stress Wave Propagation via Air-Coupled Acoustic Emission Sensors}

\author{Joshua C. Rivey, Gil-Yong Lee, Jinkyu Yang}

\address{University of Washington, Seattle, Washington, 98195, USA}

\author{Youngkey Kim}

\address{SM Instruments, Daejon, Republic of Korea}

\author{Sungchan Kim}

\address{Korea Aerospace Research Institute, Daejon, Republic of Korea}

\vspace{10pt}
\begin{indented}
\item[]February 2016
\end{indented}

\begin{abstract}

We experimentally demonstrate the feasibility of visualizing stress waves propagating in plates using air-coupled acoustic emission sensors.  Specifically, we employ a device that embeds arrays of microphones around an optical lens in a helical pattern. By implementing a beamforming technique, this remote sensing system allows us to record wave propagation events in situ via a single-shot and full-field measurement.  This is a significant improvement over the conventional wave propagation tracking approaches based on laser doppler vibrometry or digital image correlation techniques. In this paper, we focus on demonstrating the feasibility and efficacy of this air-coupled acoustic emission technique using large metallic plates exposed to external impacts.  The visualization results of stress wave propagation will be shown under various impact scenarios.  Such wave visualization capability is of tremendous importance from a structural health monitoring and nondestructive evaluation (SHM/NDE) standpoint.  The proposed technique can be used to characterize and localize damage by detecting the attenuation, reflection, and scattering of stress waves that occurs at damage locations.  This can ultimately lead to the development of new SHM/NDE methods for identifying hidden cracks or delaminations in metallic or composite plate structures simultaneously negating the need for mounted contact sensors.\\

\end{abstract}

%
%
%
%
%

\section{Introduction}

In aerospace, civil, and mechanical engineering applications, it is imperative to ensure structural integrity through the detection and characterization of damage and defects.  The presence of defects, such as cracks, dents, corrosion, delaminations, or numerous other forms of damage, can significantly reduce the inherit properties and performance of a structure, thereby increasing the chance of premature failure. Therefore, structural health monitoring (SHM) and nondestructive evaluation (NDE) have been subjects of intense studies in recent decades. In particular, increases in the use of advanced materials, sensors/actuators, and manufacturing processes has spurred the development of numerous SHM/NDE techniques. These methods include -- but are not limited to -- thermography, shearography, X-radiography, eddy current, ultrasonic C-scan, scanning laser Doppler vibrometry, and guided wave-based ultrasound techniques~\cite{Giurgiutiu, Johnson_SHM, Civil_SHM, ASNDT, Djordjevic}. They are based on thermal, electromagnetic, acoustic/mechanical, and other multi-physical feedback, and each technique offers unique advantages and shortcomings.  



Using acoustic emissions for purposes of detecting and localizing damage has been one of the widely adopted methods in the SHM/NDE community~\cite{NSF, Rizzo, Hellier}. Acoustic emissions are manifest when a material is subject to extreme stress conditions due to external loads, such that a local point source within the material suddenly releases irreversible energy in the form of stress waves. The released stress waves are transmitted to the surface of the material and then propagate outwards from the epicenter of the release source. 
Previous studies on acoustic emission have focused mostly on the onset of such acoustic emissions to locate their release sources~\cite{McLaskey, Grosse, Olson, Sengupta}. However, more useful can be that acoustic emissions are also attenuated, scattered, or reflected by discontinuities present in the material \cite{Hellier}. We identify that it is this particular property of acoustic emissions that can be exploited in order to detect and localize pre-existing damage in an inspection medium.

In this study, we experimentally demonstrate the feasibility of visualizing stress waves in an aluminum plate using acoustic emissions.  The focus is whether acoustic emission techniques can capture the scattering of stress waves due to the presence of damage resulting from an applied impact on the plate.  For visualizing such transient events, we employ a device -- referred to as an acoustic or sound camera -- that embeds an optical lens at the center and arrays of microphones around the optical lens in a helical pattern~\cite{SMI}.  Note that arrays of microphones have been used in previous studies to identify the locations of vibration sources~\cite{Dougherty}, but it has not been thoroughly explored yet to visualize stress waves in structures via air-coupled microphone sensors to the best of the authors' knowledge.  This is because of a short characteristic time -- in the order of micro-seconds -- of stress waves propagating in solids and also due to the difficulty in capturing and visualizing the wavefronts of these stress waves. 

To address the challenges associated with stress wave visualization, we implement time-domain delay-sum beamforming techniques~\cite{Mucci} based on acoustic emission information collected from the arrays of microphones.  To enhance the accuracy of the diagnostic scheme, we conduct parametric studies on various post-processing conditions, including the temporal resolution of the sensor data and the spatial resolution of the inspection plate.  Finally, the developed technique is evaluated for SHM/NDE purposes with capabilities assessed for detecting the defect location simulated with a mass placed in the path of the wave propagation on the inspection plate.  

The acoustic emission beamforming technique coupled with the sound camera device has unique advantages compared to the conventional techniques such as ultrasonic or laser based testing methods \cite{NSF}.  Many of the conventional testing methods are founded on the principle of imparting external stimuli or excitations on the inspection material and measuring differences in the received signal in order to detect whether damage is present and where it is located.  While such methods generally provide highly accurate results, they often involve slow and expensive processes to operate equipment. Conversely, acoustic emission beamforming methods can be conducted in situ and in real time, without necessitating permanently mounted contact sensors or baseline data.  Recently, laser Doppler vibrometry has gained significant attention as a means to visualize stress waves in solids and structures with an unprecedented resolution~\cite{LDV, Feng2013, Yang2014}.  However, this method requires synchronization and reconstruction of data measured from every single discretized spatial point of an inspection medium.  This is not practical given the difficulties in exciting structures in a repeated, identical fashion. Digital image correlation techniques can also visualize extremely dynamic motions, but they require speckle patterns on specimens and their field of view can be often narrow for recording high speed events~\cite{Sutton}. In contrast, the proposed acoustic emission beamforming technique is capable of conducting non-contact -- yet full-field -- visualization of inspection medium in a single shot measurement.  Consequently, we envision that this method can open new avenues to diagnosing the existence of damage in structures in a time- and cost-efficient manner by conducting simple tests.

The contents of this manuscript will address the following topics.  Section 2 contains an introduction to the time-domain delay-sum beamforming theory and a discussion of how this concept is incorporated into the study contained herein.  Section 3 describes the experimental setup used for monitoring inspection plates and tracking stress waves using the sound camera.  Section 4 provides an analysis of parametric studies performed on pristine plates that were used to determine temporospatial resolutions necessary for capabilities-centric analyses.  Finally, Section 5 concludes the study with an analysis of the acoustic emission beamforming method for identifying varying impact locations and detecting simulated damage on the inspection plate.  

%

\section{Theoretical background}

Traditional methods of source localization via air-coupled acoustic emission rely on the assumption that the recording device is pointed at or aimed in the general vicinity of the emission origin.  Acoustic beamforming is an attractive alternative for use in acoustic feedback NDE as it provides the opportunity to detect the direction of unknown acoustic emissions associated with failure events across a large spatial domain~\cite{Chen}.  While acoustic beamforming is a relatively new concept for applications related to damage localization, beamforming using microphone arrays is a standard practice for spatial isolation of sound sources.  The beamforming method -- also known as microphone antenna, phased array of microphones, acoustic telescope, or acoustic camera -- is used extensively for localizing sounds on moving objects and to filter out background noise in acoustically active environments with stationary sound sources~\cite{Michel}.  

\begin{figure}[H]
\centering
\includegraphics[width=7.25\textwidth]{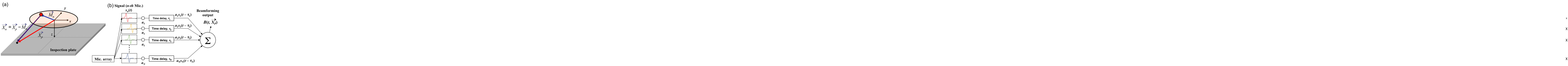}\\
\caption[Calculation of delays for beamforming algorithm]{Illustrations showing (a) location vectors of reference point, microphones, and inspection points; and (b) concept of calculating delays for a specific microphone for a given spatial position.\label{Fig:BeamformingDelays}}
\end{figure}

Figure \ref{Fig:BeamformingDelays} gives a visual representation of the basic delay-sum beamforming method.  This process can be expressed mathematically by Eq. (\ref{Eq:ConventionalBF}) in terms of the time-domain delay-sum beamforming output \cite{Mucci}:
\begin{equation}
B(t,\overrightarrow{X}_{P}) = \sum_{n=1}^{N} a_n s_n [ t-\tau_n(\overrightarrow{X}_{P}) ]
\label{Eq:ConventionalBF}
\end{equation}
where $N$ is the total number of microphones, \textit{t} is time, and $\overrightarrow{X}_{P}$ represents the position onto inspection plate with respect to the reference position (e.g., center of the microphone arrays as shown in the figure).  $a_{n}$ is a spatial shading or weight coefficient that can be applied to the individual microphones to control mainlobe width and sidelobe levels \cite{Mucci}.  In many instances, the weighting coefficients are set to unity or equal to one.  $s_n(t)$ represents the acoustic emission received by the $n$-th microphone emitted by an arbitrary sound source.

After receiving the signal, a specified delay, $\tau_n$, is imposed to the signal for each individual microphone based on the spatial domain.  Figure \ref{Fig:BeamformingDelays} shows the operation used for calculating the microphone delays. First, the spatial domain of interest is discretized into ``pixels" rendering a superimposed grid onto the spatial domain.  Second, for each point on the spatial domain the position vector $\overrightarrow{X}_{p}$ from the predetermined reference point is calculated.  Then, the position vector of that same spatial pixel from the $n$-th microphone whose position is given by $\overrightarrow{M}_{n}$ is determined as $\overrightarrow{X}_{n} = \overrightarrow{X}_{p}$ - $\overrightarrow{M}_{n}$. Finally, the difference of flight time of the acoustic signal between the two vectors is found by calculating the difference in vector magnitudes and dividing the speed of sound $c$.  That is, the time delay for each point on the inspection plate and each $n$-th microphone is defined by:  
\begin{equation}
\tau_n(\overrightarrow{X}_{P})=\frac{1}{c}\left(\Big|\overrightarrow{X}_{p}\Big|-\Big|\overrightarrow{X}_{n}\Big|\right).
\label{Eq:Time_delay}
\end{equation}


The scanning algorithm of the microphone array will perform the delay calculation operation for the entirety of the spatial domain which it is monitoring.  After the time delays for all microphones are imposed on the signal for a given spatial pixel, the transformed signal from each microphone is summed as shown in Figure \ref{Fig:BeamformingDelays}b.  The location of the sound source is determined by the delays for a specific spatial location which produced the maximum beam output, $B(t)$.  It is important to note that the microphone delays are time independent and therefore will remain constant for a given microphone to a given spatial pixel throughout the monitoring period assuming the geometry of the test setup remains constant.  While Figure \ref{Fig:BeamformingDelays} and Equation \ref{Eq:ConventionalBF} illustrate the basic delay-sum beamformer in the time-domain, there exist many variations and modifications to this algorithm, see reference~\cite{Mucci}.

\section{Experimental setup}



\begin{figure}[H]
\centering
\includegraphics[width=1.05\linewidth]{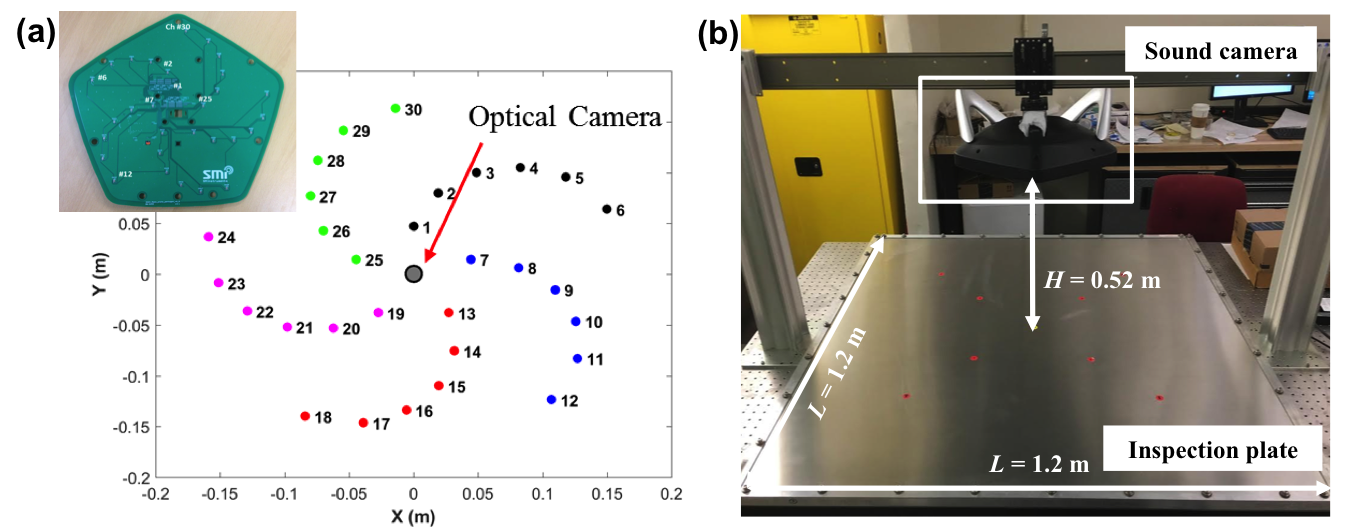}\\
\caption{(a) Arrangement of 30 MEMS microphones and an optical camera. Inset shows a digital image of the sound camera. (b) Proposed experimental setup for inducing and tracking transient waves in an aluminum plate using the sound camera.\label{Fig:ExSetup}}
\end{figure}

For the study presented herein, we used a commercial acoustic emission sensing device equipped with microphone arrays and a motion camera (SeeSV-S205 Sound Camera, SM Instruments)~\cite{SMI}.  Specifically, the sensing device consisted of a high resolution optical camera with a sampling rate of 25 frames per second (FPS) located in the center of the device.  The optical camera is surrounded by 30 high sensitivity digital micro-electric mechanical system (MEMS) microphones with a sampling frequency of 25.6 kHz arranged in five helical patterns of six microphones each as shown in Figure \ref{Fig:ExSetup}a. The figure inset shows a digital image of embedded microphones along with the optical lens.



Figure \ref{Fig:ExSetup}b shows the experimental setup used in this study to induce and track the transient waves in the aluminum plate.  The images show the 1.2 m $\times$ 1.2 m $\times$ 1.02 mm 6061-T6 aluminum plate mounted to an optical table using a rail system to create a fixed boundary condition around the plate. 
The plate was secured between the angle bars and the square tube with fasteners placed every six inches that went through all pieces to effectively fix the boundary of the plate and suspend it above the optical table.

     
Once the plate was installed, the acoustic camera was mounted above the center of the plate at a height of 0.52 m as shown in Figure \ref{Fig:ExSetup}b.  
Impacts were introduced into the plate via manually tapping the plate using the tip of a hexagonal wrench while simultaneously capturing the acoustic signal recorded by the 30 microphones.  Afterwards, post-processing was performed to calculate the time delays and beamforming output based on Equations (\ref{Eq:ConventionalBF}) and (\ref{Eq:Time_delay}) in order to produce the acoustic images.



\section{Parametric studies on pristine plates}

In this study different parameters were systematically varied to investigate and assess the wave propagation tracking capabilities of the sound camera in addition to determining post-processing parameters used in subsequent analyses. These parameters included the temporal and spatial resolutions in post-processing. This section discusses the results from these parametric studies given pristine plates. For all images shown of the wave propagation tracking, the area presented in the image represents that of the entire inspection plate.


\subsection{Effects of temporal resolution}

The temporal resolution was investigated with the goal of achieving a smooth propagation of the transient wave front.  In this study, the sound camera MEMS microphones imposed a hardware limitation on the sampling frequency restricting it to 25.6 kHz.  This meant the time between individual samples was slightly greater than 39 $\mu$s.  During investigation of wave propagation velocities, the speed of major flexural waves propagating in the 1.02 mm thick plate was approximately 1,700 m/s (to be further discussed below).  At this wave speed and the given sampling frequency, the wave front propagated approximately 66.4 mm between each sample or 5.4$\%$ of the total plate width, resulting in a very coarse propagation tracking ability. 

\begin{figure}[H]   
\centering
\includegraphics[width=1\linewidth]{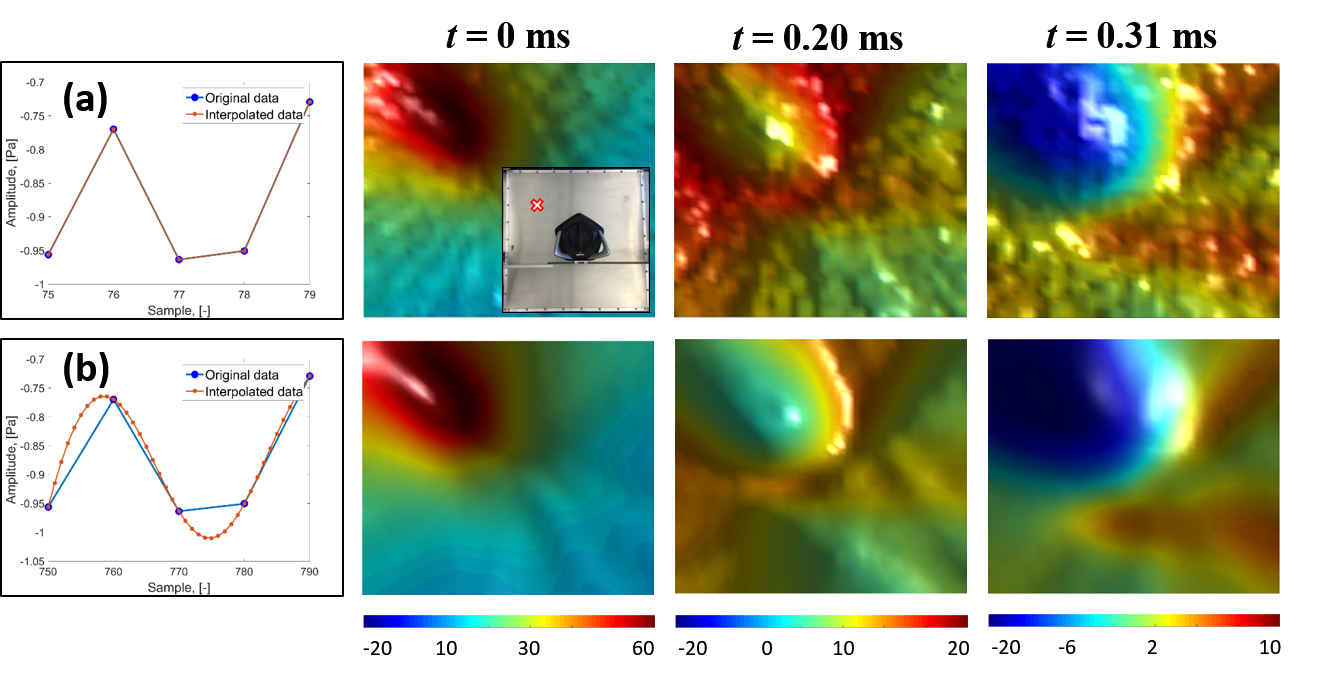}\\
\caption[Parametric study - Analysis of temporal resolution]{Effects of changing temporal resolution using interpolation of results in frequency domain.(a) Original temporal data and the post-processed surface maps showing stress wave propagation at $t$ = 0, 0.20, and 0.31 ms. (b) Interpolated temporal data and the post-processed surface maps showing stress wave propagation at $t$ = 0, 0.20, and 0.31 ms. For all surface maps, the spatial resolution was fixed at 30 mm between adjacent pixels.  The colorbar at the bottom shows the intensity of the beamforming output in Pa.\label{Fig:TemporalResolution}}
\end{figure} 

To reduce the effective propagation distance between sequential samples and to improve the resolution of the wave propagation tracking, an interpolation of the raw acoustic signal was performed.  Figure \ref{Fig:TemporalResolution}a shows a window of the raw signal with no interpolation applied (left panel), in which we observe the drastic amplitude changes of pressure measured from a microphone between subsequent samples.  The beamforming results of the stress wave propagation for the raw signal without interpolation are shown in the three images to the right of the signal.  The inset image at $t$ = 0 ms in Figure \ref{Fig:TemporalResolution}a represents the top left impact induced in the inspection plate.  

Figure \ref{Fig:TemporalResolution}b shows the same raw signal as the one above it, but with a frequency-domain fast fourier transformation interpolation applied to reconstruct the signal.  An upsample factor of 10 was chosen for the interpolation in order to decrease the time between samples to 3.9 $\mu$s.  This meant that the propagation distance between samples was reduced to 6.6 mm and only 0.5$\%$ of the total plate width. 
The drastic effects of the interpolation are seen in the series of images to the right of the second signal.

Comparing the two sets of images presented in Figure \ref{Fig:TemporalResolution}, we find that both approaches successfully identify the impact location with a reasonably high accuracy (to be further discussed in the later section). However, the results from the raw signal do not show clear boundaries of wave front while the second set of images associated with the interpolated data show the increased definition of the wave front.  Additionally, a significant amount of acoustic background noise was removed in the case of interpolation, which resulted in a refined beamforming image.  Looking at $t$ = 0.20 ms in Figure \ref{Fig:TemporalResolution}, it is clear that the wave front amplitude is reduced in the reconstructed signal image.  The decreased amplitude is due to the more accurate representation of the wave front shape resulting from the interpolation of the acoustic signal.  This reduces exaggerated pressure amplitude changes observed from the original acoustic signal.

While the resolution of the wave propagation was achieved by increasing the upsample factor, a computational penalty was incurred.  Based on parametric studies on different upsample factors, the upsample-computational time relationship was approximately linear with an increase in computational time. Specifically, for the upsample factor of $n$, the beamforming computational time is increased $(0.1044 \pm 0.014) \times n$ times compared to the original raw signal.  The reduced computational efficiency was deemed an acceptable cost for the increased resolution gained using the reconstructed signal and necessary for identifying and localizing masses. For all subsequent simulations, we used an upsample factor of 10.  

\subsection{Effects of spatial resolution}

\begin{figure}[H]
\centering
\includegraphics[width=1\linewidth]{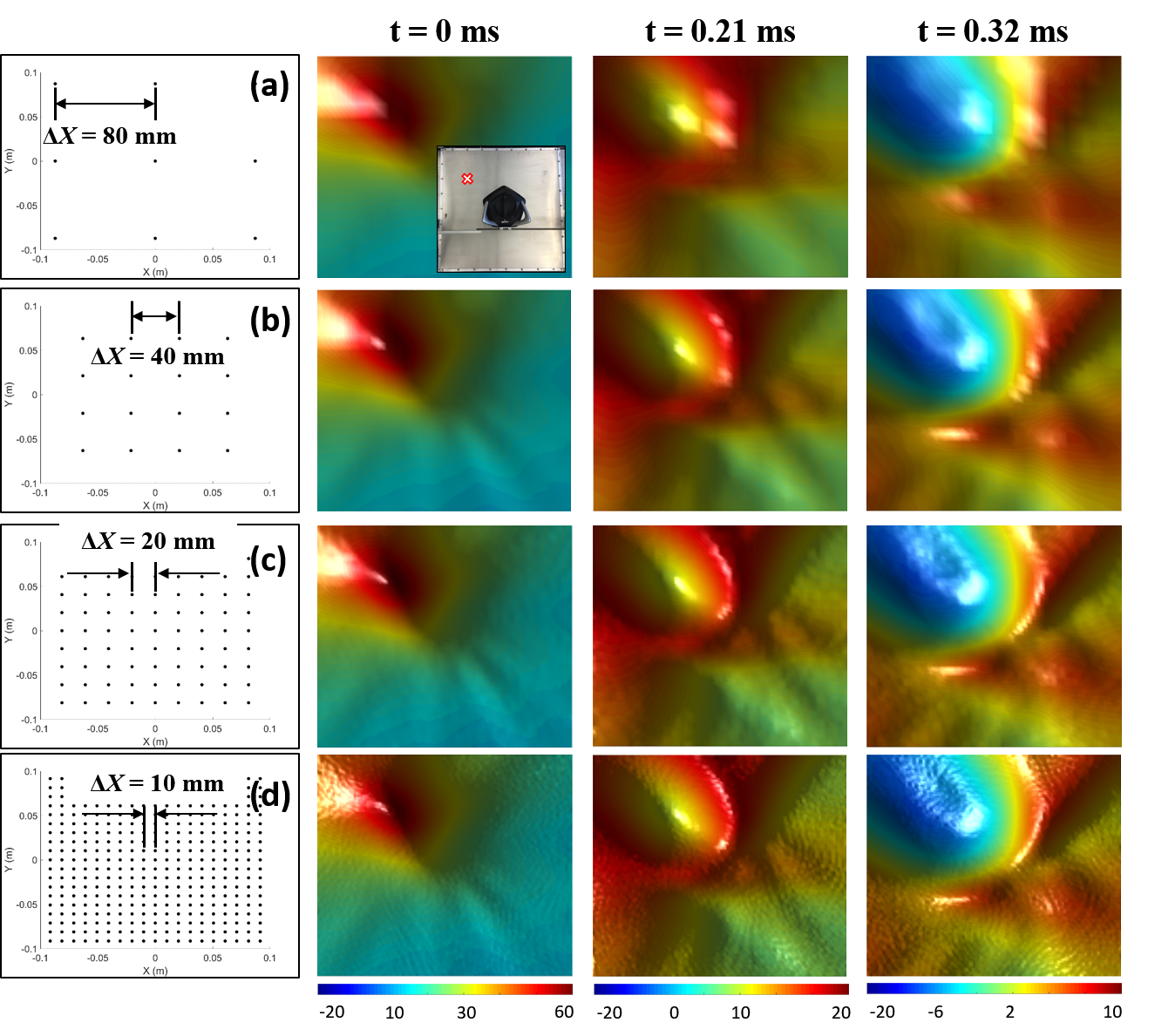}\\
\caption[Parametric study - Analysis of spatial resolution]{Effects of changing spatial resolution by decreasing step size between ``pixels" on virtual inspection plane. Left column shows spatial resolution, while the next three right columns show surface maps of wave propagation for $t$ = 0, 0.21, and 0.31 ms under the spatial resolution of (a) $\Delta$X = 80 mm; (b) 40 mm; (c) 20 mm; and (d) 10 mm.
\label{Fig:SpatialResolution}}
\end{figure}

Now we investigate the effect of spatial resolution in an attempt to further refine the wave front throughout the propagation time.  Figure \ref{Fig:SpatialResolution} shows the results of the spatial resolution study. Here the spatial resolution was systematically halved for each simulation throughout the parametric study.  Figure \ref{Fig:SpatialResolution}a and subsequent beamforming images show the results for a spatial resolution ($\Delta$X) of 80 mm between pixels.  While the impact point in the surface map is well localized (compare with the actual impact location as shown in the inset image), the wave front becomes blurry as it begins to propagate into the far-field as seen at times $t$ = 0.21 ms and $t$ = 0.32 ms.  

Figures \ref{Fig:SpatialResolution}b-c and associated images show spatial resolutions of 40 mm and 20 mm between pixels, respectively.  Both sets of images show an accurate impact localization and increased definition of the wave front as the spatial resolution increases.  Figure \ref{Fig:SpatialResolution}d and corresponding wave propagation images represent the case of $\Delta$X = 10 mm between pixels. 
Compared to all previous results, these images show a very definitive wave front at $t$ = 0.21 ms.  The image at $t$ = 0.32 ms for a spatial resolution of 10 mm solidifies that increased spatial resolution can maintain the wave front definition into the far-field of the plate while post-impact saturation behind the wave front clearly emanates from the impact location.

While a 10 mm displacement between adjacent pixels may be excessive by creating ripples, it increases the chance of detecting an artificially created damage (to be discussed in Section 5) and accurately determining its location.  However for each halving of the spatial resolution, the computational time is increased by a factor of approximately 4.25, thereby resulting in an approximately power-law relationship equal to $(2.08 \pm 0.042)$ increase in computational time as the spatial resolution is increased (that is $\Delta$X decreases).  The power-law relationship is due to the spatial resolution affecting both dimensions of the inspection plate meaning that the increase in spatial resolution is squared with each iteration.  While this leads to a significant increase in  computational time, the increase was again considered as an acceptable trade-off for the gained wave propagation resolution. This would prove necessary for detecting and localizing artificially created damage on the plate in subsequent analyses.  Following the temporal and spatial resolution studies, a temporal upsample factor of 10 and a spatial resolution of 10 mm were selected for use in all following analyses.  

\section{Feasibility studies on the applications to NDE/SHM}

In this section, we assess the feasibility of using the beamforming-based sound camera technique for (i) identifying impact location for real-time SHM applications and (ii) detecting artificial damage location for potential NDE applications. 

\subsection{Identification of impact location}

\begin{figure}[H]
\centering
\includegraphics[width=1\linewidth]{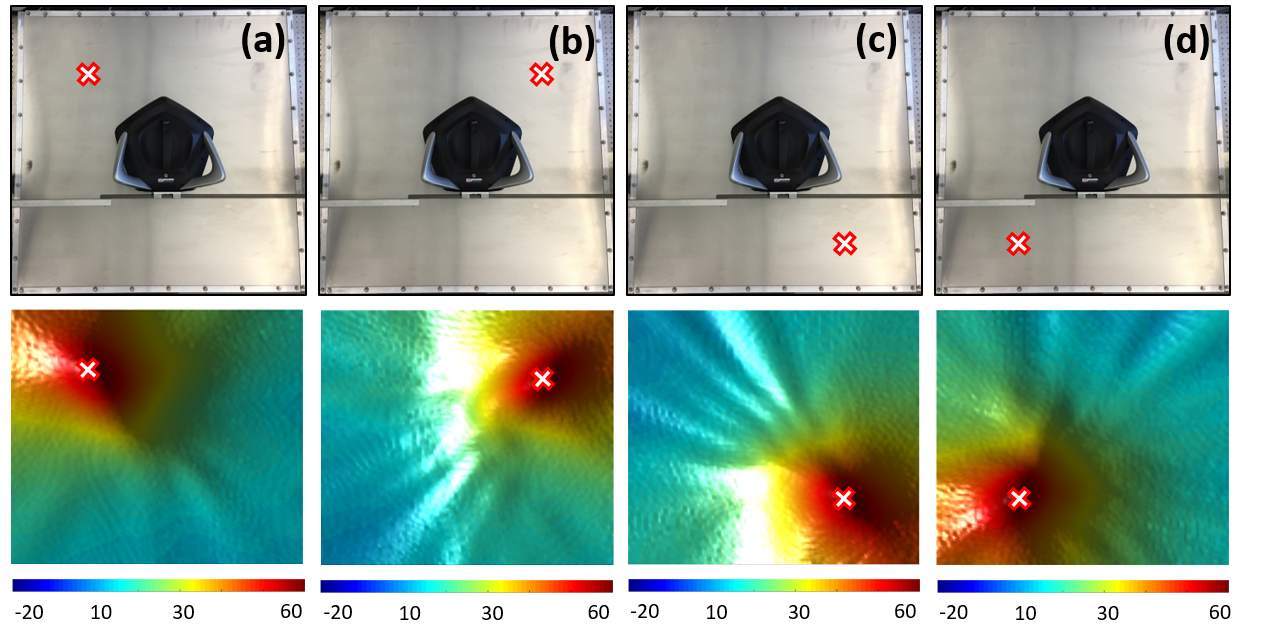}\\
\caption[Parametric study - Analysis of impact location]{Impact locations (top row) and post-processing results (bottom row) showing impact detection capability of the sound camera and beamforming algorithm. Location of impact was varied in four cases: (a) top left, (b) top right, (c) bottom right, and (d) bottom left. \label{Fig:ImpactLocation}}
\end{figure}

Once simulation parameters were determined, tests were performed to characterize detection and localization capabilities of the sound camera.  Prior to masses being placed on the plate, it was necessary to determine if different impact locations could be identified since a single impact location in the top left corner of the plate had been used for all previous parametric studies.  Additionally, these tests would serve as the first quantitative indication of the detection and localization capabilities.  Figure \ref{Fig:ImpactLocation} shows the qualitative results of the impact location study. 

Figure \ref{Fig:ImpactLocation}a-d show the beamforming images at the moment of impact for impacts located in the four corners of the plate.  With the center of the inspection plate above which the camera is suspended defined as the origin of the plate, the actual measured impact locations are given in Table \ref{Tab:ImpactLocation}.  For each of the impact locations shown, the images suggest that the impact location is properly identified and localized using the sound camera and beamforming algorithm.  The images shown were used to identify the point of maximum amplitude at the instant of impact which was defined as the impact location.  These identified locations are given as the identified locations in Table \ref{Tab:ImpactLocation}.  

\begin{table}
\caption[Quantification of impact localization]{Quantification of impact localization.\label{Tab:ImpactLocation}}
\begin{indented}
\item[]\begin{tabular}{@{}lllll}
\br
\vtop{\hbox{\strut Impact}\hbox{\strut Location}} & \vtop{\hbox{\strut Actual}\hbox{\strut Position [m,m]}} & \vtop{\hbox{\strut Identified}\hbox{\strut Position [m,m]}} &\vtop{\hbox{\strut Absolute}\hbox{\strut Difference [m]}} & \vtop{\hbox{\strut Normalized}\hbox{\strut Error [Num. Pixels]}}\\
\mr
Top Left	 & (-0.3032,0.3048)  & (-0.2743,0.3251)  & 0.0353 & 3.53\\ 
Top Right	 & (0.3048,0.3064)   & (0.3556,0.3150)   & 0.0515 & 5.15\\ 	
Bottom Right & (0.3056,-0.3064)  & (0.3150,-0.3251)  & 0.0209 & 2.09\\ 
Bottom Left	 & (-0.3040,-0.3056) & (-0.2946,-0.2946) & 0.0145 & 1.45\\   
\br
\end{tabular}
\end{indented}
\end{table}

We quantify the percent error between the actual and identified impact locations as below: 
\begin{equation}
Normalized Error = \frac{\sqrt{\left({X_{actual} - X_{identified}}\right)^2+\left({Y_{actual} - Y_{identified}}\right)^2}}{\Delta X},
\label{Eq:ImpactError}
\end{equation}
where $\Delta X$ represents the pixel size. Looking at actual and identified positions in Table \ref{Tab:ImpactLocation} (i.e., ($X_{actual}, Y_{actual}$) and ($X_{identified}, Y_{identified}$)), it is observed that all impacts were located relatively accurately with regards to identified positions being in the general vicinity of the actual known position of the impact.  Using Equation \ref{Eq:ImpactError}, the normalized error in terms of number of pixels between the actual and identified impact locations with respect to the spatial discretization was calculated.  The calculated errors are given in Table \ref{Tab:ImpactLocation}.  Given the $\Delta$X of 10 mm used for this study, the normalized error between the actual and identified impact locations represent differences ranging from 14.5 mm to 51.5 mm as shown in Table \ref{Tab:ImpactLocation}.  The largest error was calculated for the top right impact location and found to be 5.15 pixels or approximately 4.22$\%$ of the plate width. This means that, with respect to the total spatial domain in this case, the localization was still relatively accurate.  The larger errors could be attributed to not a fine enough spatial resolution which leads to a larger error between the actual and identified locations.  Additionally, the larger error values could be due to human error when initiating the impacts on the plate.  Impacts were incited by manually tapping the inspection plate at predetermined locations which are noted as the actual impact locations.  However, if the impact on the plate were not located at precisely these locations, this would greatly affect the error values as the identified impact location is based off experimental data which assumes the impact is imparted at the actual location.  Despite some seemingly large error values between actual and identified impact locations this study indicates the sound camera and beamforming algorithm were able to fairly accurately localize different impact locations and track the resultant transient wave across the plate. 

\subsection{Identification of pre-existing artificial damage}

\begin{figure}[H]
\centering
\includegraphics[width=1\linewidth]{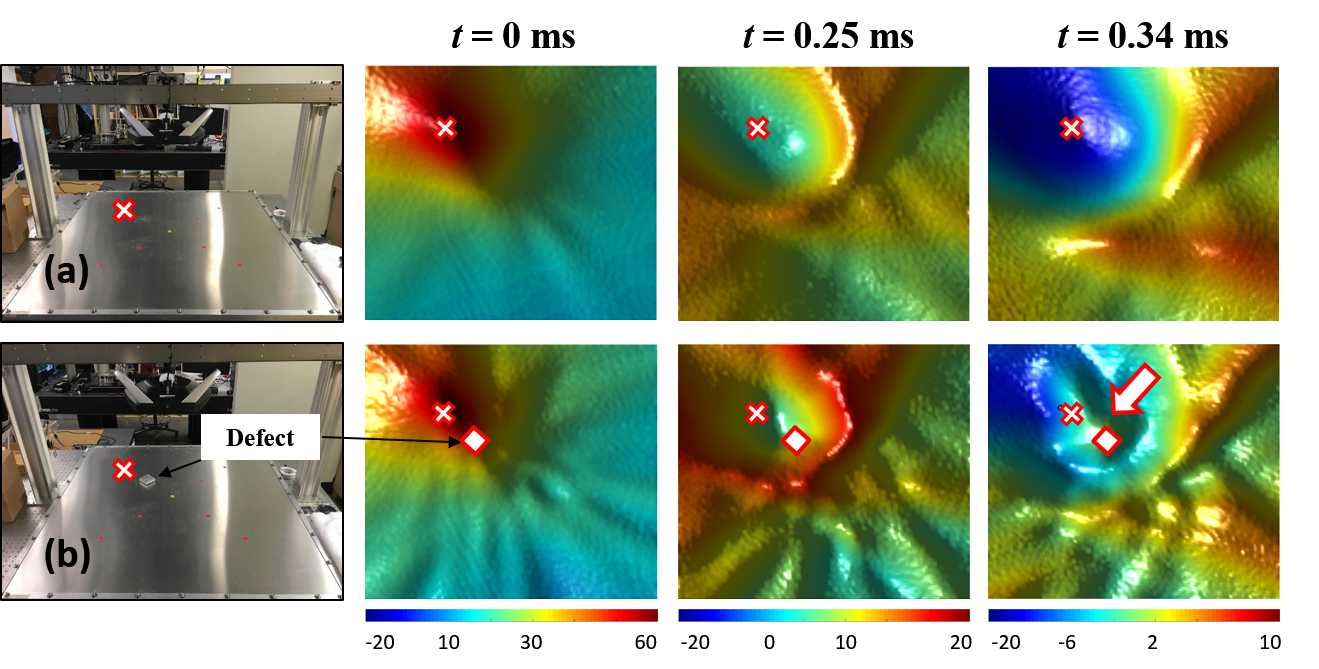}\\
\caption{Beamforming results for plate without and with mass.  Results suggest that sound camera using acoustic beamforming can be used for detection of discontinuities.\label{Fig:DamageDetection}}
\end{figure}

Once it was established that the sound camera could detect and sufficiently track the transient wave, the impact location was fixed and masses were added to the plate to determine the capability of the sound camera and beamforming algorithm to detect the discontinuities which represented psuedo-damage cases.  The impact location used is a top left impact as shown in Figure \ref{Fig:DamageDetection}a-b.  The mass used for this study was a 0.615 kg, 0.0635 m $\times$ 0.0635 m $\times$ 0.0196 m stainless steel block.  The mass was affixed to the inspection plate using silicon sealant tape that was applied to the entire contact surface of the mass.   The use of masses to simulate damage has been used on many occasions to assess the capabilities of different SHM/NDE techniques~\cite{Kody, Adams}. Figure \ref{Fig:DamageDetection}b shows the mass was place halfway between the impact location and the center of the plate meaning the mass was approximately 0.215 m from the impact location.  Figure \ref{Fig:DamageDetection} shows the wave propagation results for the inspection plate without and with mass present.

Figure \ref{Fig:DamageDetection} uses a red cross and a red square to mark the approximate location of the impact and mass, respectively, on the plate.  At $t$ = 0 ms, the instant of impact is shown for both the without- and with-mass cases.  In these images look very similar between the two cases and there is no indication of a mass being present on the plate.  The images at $t$ = 0.25 ms show the wave front just after it has propagated past the location of the mass.  The first difference is seen in the wave front definition.  While the no-mass case maintains a smooth and symmetric wave front, the mass case displays a more turbulent wave front with a flatten of the wave front on the far side of the mass relative to the impact point.  Additionally, the widespread saturation behind the wave front seen in the no-mass case is not seen in the with-mass case.  In the with-mass case, the area behind the wave front maintains a higher acoustic level throughout compared to the no-mass case contributing to the less defined wave front.  Finally, there is much more acoustic noise in the farfield of the plate for the with-mass case compared to the no-mass case where the farfield acoustic level is approximately equal to the behind the wave front at $t$ = 0.25 ms. 

\begin{table}
\caption[Quantification of mass detection and localization capability]{Quantification of mass detection and localization capability.\label{Tab:DamageDetection}}
\begin{indented}
\item[]\begin{tabular}{@{}lllll}
\br
\vtop{\hbox{\strut Mass}\hbox{\strut Location}} & \vtop{\hbox{\strut Actual}\hbox{\strut Position [m,m]}} & \vtop{\hbox{\strut Identified}\hbox{\strut Position [m,m]}} &\vtop{\hbox{\strut Absolute}\hbox{\strut Difference [m]}} & \vtop{\hbox{\strut Normalized}\hbox{\strut Error [Num. Pixels]}}\\ 
\mr
Point 1	 & (-0.1524,0.1524)  & (-0.1219,0.193) & 0.0508 & 5.08\\ 
\br
\end{tabular}
\end{indented}
\end{table}

At $t$ = 0.34 ms, the most noticeable difference between the no-mass and with-mass cases is the high amplitude acoustic event present at the location of the mass in the second case.  For the no-mass case, the wave front is losing amplitude and definition as it propagates and dissipates into the farfield and the area behind the wave front is becoming saturated with low amplitude acoustic levels.  However, the backside of the wave front remains very symmetric with respect to the impact location and there is very little excessive acoustic features seen on the plate.  In the with-mass case, the overall wave front condition appears very similar to that for the no-mass case.  In terms of dissipation and primary acoustic features, the wave fronts for the two cases display very similar amplitudes and structures at $t$ = 0.34 ms.  However, despite the similarities the with-mass case wave front retains a much less refined wave front at this instance with much more acoustic noise seen in the farfield of the plate.  As aforementioned, the most observable and desirable difference is the presence of the acoustic event emanating from the mass location denoted by the red arrow in Figure \ref{Fig:DamageDetection} which alludes to the presence of a discontinuity at this location on the inspection plate.

Similar to the impact localization, the maximum amplitude of the acoustic feature near the known location of the mass was used as the identifying location of the mass.  This was then compared to the actual location of the mass given in Table \ref{Tab:DamageDetection} and using Equation \ref{Eq:ImpactError} the normalized error was calculated between the actual and identified position with respect to the spatial discretization.  The calculated error of only 5 spatial pixels represents a fairly high degree of accuracy when identifying the known position of the mass.  Given that the mass is 63.5 mm $\times$ 63.5 mm, the mass itself is 6.35 pixels $\times$ 6.35 pixels.  While the exact coordinates of the actual and identified positions do not necessarily correlate, it is likely affected by the acoustic feedback from the mass not emanating from the center of the block.  Therefore, if the maximum amplitude were located at one of the points of the mass as it seems to be in Figure \ref{Fig:DamageDetection} (corner nearest top edge of plate), this alone would considerably affect the error value.  Given the normalized dimensions of the mass, the identified position of the acoustic emission denoting the location of the mass is likely within the bounds of the mass despite the relatively large normalized error.
\section{Conclusions}

This study demonstrated the visualization of stress wave propagation across an aluminum plate using an air-couple microphone array.  The scope of the beamforming method for visualizing stress waves was investigated through various parametric studies that resulted in the establishment of post-processing parameters of upsample factor and spatial resolution for further testing.  Subsequent testing revealed the ability of the sound camera to accurately identify various impact locations on the inspection plate.  Finally, the beamforming method was used to detect and localize a mass on the inspection plate used to simulate the presence of damage.  The sound camera and derived beamforming method were able to relatively accurately locate the position of the mass through the detection of acoustic emission structures indicating the existence of a discontinuity on the inspection plate.  This study showed the potential for an acoustic emission beamforming based method to be used for damage detection in SHM/NDE applications. While this study focused on the feasibility of the proposed technique, further studies need to be conducted to find more sophisticated beamforming techniques and to optimize their post-processing parameters. Corresponding experiments will be also conducted in comparison to other experimental techniques, such as laser Doppler vibrometry and digital image correlation techniques. Lastly, the authors also plan to perform numerical studies of air-coupled acoustic emission events using finite element analysis. 

\section*{Acknowledgments}

This research was supported by INNOPOLIS Foundation grant funded by the Korean government (Ministry of Science, ICT $\&$ Future Planning, Grant number: 14DDI084) through SM Instruments Inc.  We also acknowledge the research grant from the Joint Center for Aerospace Technology Innovation (JCATI) from Washington State in the USA.  We thank researchers at SM Instruments, including InKwon Kim and JunGoo Kang, for their technical assistance.  


\end{document}